\UseRawInputEncoding
\documentclass{article}

\usepackage{PRIMEarxiv}

\usepackage[utf8]{inputenc} 
\usepackage[T1]{fontenc}    
\usepackage{booktabs}       
\usepackage{amsfonts}       
\usepackage{nicefrac}       
\usepackage{microtype}      
\usepackage{fancyhdr}       
\usepackage{graphicx}       
\graphicspath{{media/}}     
\usepackage{float}

\title{RESEARCH on the accumulation effect model of technological innovation in textile industry based on chaos theory}

\author{
  Xiangtai Zuo\\
  Accounting College \\
  Wuhan Textile University \\
  Wuhan, China\\
}

\begin{document}
\maketitle

\begin{abstract}
Technological innovation is one of the most important variables in the evolution of the textile industry system. As the innovation process changes, so does the degree of technological diffusion and the state of competitive equilibrium in the textile industry system, and this leads to fluctuations in the economic growth of the industry system. The fluctuations resulting from the role of innovation are complex, irregular and imperfectly cyclical. The study of the chaos model of the accumulation of innovation in the evolution of the textile industry can help to provide theoretical guidance for technological innovation in the textile industry, and can help to provide suggestions for the interaction between the government and the textile enterprises themselves. It is found that reasonable government regulation parameters contribute to the accelerated accumulation of innovation in the textile industry.
\end{abstract}


\section{Introduction}
\begin{figure}[H]
    \centering
    \includegraphics[scale=0.7]{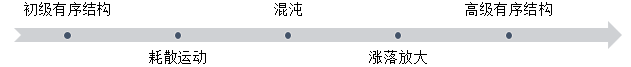}
    \caption{System evolution illustration.}
    \label{fig:fig1}
\end{figure}

Hamel and Prahalad \cite{2005Competing} point out that the rapid changes in technology and the non-stationary nature of market demand create dynamic behavioural characteristics of the environment, and that in today's society, where technological innovations are proliferating and transformations are taking place rapidly, the cumulative effects of technological innovation in firms are more likely to take on the characteristics of a complex, irregular and non-periodic non-linear dynamical system. This non-linear characteristic seems to be chaotic, but in fact it is in order. From the perspective of non-linear theory, the evolution of a system can be summarised as shown in Figure \ref{fig:fig1} . McBride \cite{2010Chaos} defines chaos as a qualitative study of the unstable, acyclic behaviour of a deterministic nonlinear dynamical system, and Sardar and Abrams \cite{2004Introducing} argue that 'order and chaos coexist, order in chaos and chaos in order. The presence of chaos in a non-linear system should be categorised and discussed. There are situations where chaos should be avoided, times when it should be controlled, and times when it should be exploited. 

So as of now, research on models of the cumulative effects of technological innovation in the textile industry is still very limited and far from systematic. The textile industry, as an indispensable part of the population, plays an important role in clothing, food, housing and transport, and has experienced a long period of development. The cumulative effect of technological innovation in the industry is also prominent, so the textile industry is chosen as the subject of this study. This paper discusses the following: 1. the application of a model of the cumulative effect of technological innovation in the textile industry from the perspective of chaos theory; and 2. the provision of countermeasures for technological innovation in the textile industry to appropriately control and manage chaos and increase the effective innovation rate of the industry.

\section{A chaotic model of the cumulative effect of technological innovation}
\subsection{Chaos theory}
The basic characteristics of chaotic motion are as follows: deterministic, non-linear, sensitive dependence on initial conditions and non-periodic. The connotation of chaos is reflected in the fact that it is a seemingly random dynamic behaviour generated by a non-linear system \cite{R1976Simple} . A deterministic system is said to be chaotic if, in the absence of external stochastic influences, 1. the state of motion of the system is irregular and complex, similar to the Brownian motion of small molecules; 2. the system has a sensitive dependence on the initial conditions; and 3. some characteristic of the system (e.g. positive Lyapunov exponent, positive topological entropy, attractor of fractional dimension, etc.) has little to do with the choice of initial conditions.
In this paper, on top of the basic framework of previous studies, a chaotic economic model of the cumulative effect of technological innovation in the textile industry is developed using the worm mouth model in chaos theory as an example, with appropriately given parameters so that it can be used to describe the dynamic evolution of the textile industry.

\subsection{The textile industry is chaotic in nature}
The evolution of the textile industry is irregular and complex. The reason is that it is impossible to predict which textile company will launch a new garment that will be liked by the people, or to judge success or failure by its industry innovations. The system of cumulative effects of technological innovation in the textile industry is sensitive to the initial conditions - the cumulative effects of the initial innovation - and is dependent on them. Moreover, Mensh \cite{1979Stalemate}, Houstein et al. \cite{1982Long}, Kleinknecht \cite{1987Are} and Silverberg et al. \cite{1993Long} have successively conducted empirical studies on the distribution function of the temporal occurrence pattern of innovation using relevant statistical data, and the results show that the realisation of innovation is a similar Poisson distribution and has an exponential growth trend of stochastic process. Therefore, both the textile industry itself and its innovation accumulation are chaotic in nature.

\subsection{A chaotic model of the cumulative effect of innovation in the textile industry}
Let $X_t$ be the marginal contribution of the accumulation of innovative technology in the textile industry system to the growth rate of the total economic volume of the textile industry system at a certain moment $t$, i.e. when the accumulation of innovation in the textile industry increases by $1\%$, the total economic volume of the textile industry increases by $X_t\%$, and call $X_t$ the accumulation effect of innovation. It is easy to see that $X_t$ is a function about time, that is, the value of $X_t$ is time-dependent. Similar to the variables used to describe the changing state of economic growth, such as capital output rate, labour productivity and capital-labour ratio, $X_t$ should also be a state variable to describe the evolution of the textile system.
The contribution of innovation to economic growth in the textile industry is much more than a simple linear accumulation of individual innovation contributors, and therefore $X_t$ is a more complex variable, as are variables such as the capital output rate, labour productivity and the capital-labour ratio. According to the definition of $X_t$, its variation mainly reflects fluctuations in the structure of productivity within the textile industry due to the continuous generation and accumulation of innovation. Silverberg \cite{1993Long} developed a model of technological progress and its evolutionary chaos, which not only explains the existence of the impact of innovation, but also shows that innovation makes productivity move in an irregular cycle, i.e. chaos. This is the theoretical basis for this paper's $X_t$.    \\
Introducing the model more commonly used in chaos economics for studying chaos in economic growth, as shown in equation \ref{equ1}.

\begin{equation}\label{equ1}
    X_{t+1} = X_{t}^{\beta} \frac{\sigma A}{1+\lambda} ( s-X_t )^{\gamma}
\end{equation}

where $X_t$ is the capital-labour ratio, $\sigma (\sigma >0)$ is the savings rate, $\lambda (0<\lambda <1)$ is the natural growth rate of labour, $A(A>0)$ is the technological progress factor, $\beta (\beta >0)$ is the elasticity of the capital-labour ratio, $\gamma$ is a constant greater than zero, and $s$ is the maximum capital-labour rate.    \\
When $\beta=\gamma=s=1$, let $\mu=\sigma A/(1+\gamma) $, then \ref{equ1} can be transformed into \ref{equ2}.

\begin{equation}\label{equ2}
    X_{t+1} = \mu X_{t}(1-X_{t})
\end{equation}

equation \ref{equ2} is the general form of the insect-population model. The purpose of introducing the transformation of equation \ref{equ1} into equation \ref{equ2} is to construct a chaotic economic model of the cumulative effect of technological innovation by analogy with a chaotic initialization \cite{2018Improving}. Drawing on the insect-population model, the following assumptions are introduced: 1) Technological innovation does not arise out of thin air, and existing innovations are usually the basis for subsequent innovations, similar to the relationship between parent and offspring insects in nature. 2) Technological innovation is measured by its level of innovation, and usually there is no mixture of parent and offspring, i.e., there is a certain amount of substitution of the firm's new product for the old one. 3) The technological innovation of an enterprise is limited by its own economic resources, just like the survival environment of a worm in nature. Therefore, this paper considers that technological innovation in textile enterprises meets the prerequisites of the insect-population model. According to the previous paper, the cumulative effect of technological innovation is a chaotic economic variable, which evolves in much the same way as the capital-labour ratio, with a non-linear evolution mechanism. Therefore, a chaotic economic model of the cumulative effect of technological innovation can be defined similarly - equation \ref{equ3}.

\begin{equation}\label{equ3}
    X_{t+1} = T \epsilon X_{t}(1-X_{t})
\end{equation}

where $X_t \in (0,1)$, $\epsilon \in (0,10)$, $T\epsilon \in (0,4)$, $X_t$ represents the proportion of the cumulative effect of technological innovation at time $t$, as the state variable of the textile system; $\epsilon$ denotes the government regulation parameter; $T$ is the specific coefficient of the textile firm (i.e. the combined coefficient of the growth rate of technological inputs $\alpha$, the proportion of technological content of output $\beta$ and the annual growth rate of labour force $n$ at a point in time, which The relationship between them is $T=\frac{\alpha + \beta}{1+n}$, where $\alpha, \beta, n \in (0,1)$, $T\epsilon$ together constitute the innovation control parameters of the textile enterprises themselves.

\subsection{Chaos of the model}
The chaotic state of \ref{equ3} can be determined using the Li-Yorke theorem. Constructing the function,

\begin{equation}\label{equ4}
    f(x)=T\epsilon x(1-x)
\end{equation}

It is not difficult to determine where $f(0)=0$ and $f(x)$ is a single-peaked function. \\
First, determine the point $x^*$ at which $f(x)$ reaches its maximum value. $x^*$ represents the maximum technological innovation accumulation effect value of $f(x)$. $x^*$ can be obtained by solving the following first order partial derivative equation.

\begin{equation}\label{equ5}
    \frac{\partial f(x^*)}{\partial x^*}=0
\end{equation}

The mapping point $x^*$ corresponding to the value of the maximum reachable function is obtained,

\begin{equation}\label{equ6}
    x^*=1/2
\end{equation}

The maximum innovation accumulation effect value $X_{max}$ is,

\begin{equation}\label{equ7}
    x_{max}=f(x^*)=\frac{T\epsilon}{2}
\end{equation}

From the above definition, it follows that the value of the maximum innovation accumulation effect should satisfy,

\begin{equation}\label{equ8}
    x_{max} \le s
\end{equation}

Next, determine the initial image point $x_i$ , and from equations \ref{equ4} and \ref{equ6}, we have,

\begin{equation}\label{equ9}
    T\epsilon x(1-x)=1/2
\end{equation}

The left end of equation \ref{equ9} is equation \ref{equ4}, a single-peaked function, so the smaller root is taken to be $x_i$, 

\begin{equation}\label{equ10}
    x_i = \frac{T\epsilon - \sqrt{(T\epsilon)^2-T\epsilon}}{2T\epsilon}
\end{equation}

Finally, determine its third-order mapping $f(x_{max})$,

\begin{equation}\label{equ11}
    f(x_{max}) = \frac{T^2\epsilon ^2}{2}(1-T\epsilon /2)
\end{equation}

provided that the first-, second- and third-order mappings $x^*$,$x_{max}$, $f(x_{max})$ generated by the initial preimage $x_i$ are mapped from the interval $[0,s]$ to its own interval $[0,s]$ and satisfy the sufficient conditions for chaos in the Li-Yorke theorem, i.e.,

\begin{equation}\label{equ12}
    0 \le f(x_{max}) \le x_i \le x^* \le x_{max}
\end{equation}

\begin{figure}[H]
    \centering
    \includegraphics[scale=0.7]{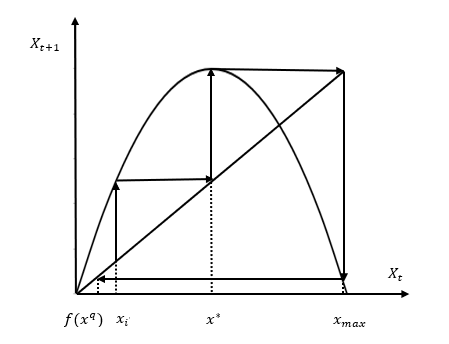}
    \caption{Chaotic phase diagrams in textile industry.}
    \label{fig:fig2}
\end{figure}

Then the textile industry system \ref{equ3} appears chaotic, as shown in Figure \ref{fig:fig2}, which is a chaotic phase diagram regarding the proportion of the cumulative effect of two adjacent generations of innovation in the textile industry innovation system. Where the curve is the image of equation (3) and the straight line is $X_{t+1} = X_t$, which represents the process of converting a preimage point to the next preimage point.

\section{Chaotic nature of the model}
\subsection{Local structural stability and cyclic bifurcation propertiest}
\begin{figure}[H]
    \centering
    \includegraphics[scale=0.7]{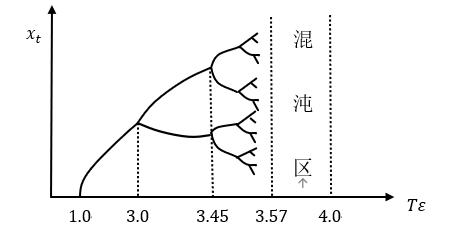}
    \caption{Bifurcations.}
    \label{fig:fig3}
\end{figure}

For model \ref{equ3}, specifically, when $T\epsilon$ has a determined value, if the corresponding $x_t$ also has only one value (indicating that the firm's innovation reaches a uniquely determined proportion, i.e. the level of intensification). That is, the technological innovation effect of a textile firm reaches a uniquely determined level. Then the period of $x_t$ is said to be $1$; if at this point $x_t$ has two values corresponding to it, then the period of $x_t$ is said to be $2$; if at this point $x_t$ has n values corresponding to it, then the period of $x_t$ is said to be $n$. In particular, the structure of $x_t$ is unstable when there is a bifurcation and the steady state can only be maintained when the period of $x_t$ is determined. This also means that a stable accumulation of technological innovation in textile companies can only be guaranteed when $T\epsilon$ takes on a value within a particular interval. In other words, a good accumulation of technological innovation in the textile industry can be maintained to a certain extent when the role of policy innovation and industry-specific coefficients are well controlled. A diagram of the iterative process for varying values of the control parameter $T\epsilon$ from $0$ to large is shown in Figure \ref{fig:fig3}.

\subsection{Effect of parameter values on the initial innovation ratio in the textile industry}
Based on its definition of firm intensification in a chaotic economic model, the effect of the $T\epsilon$ parameter on the initial innovation ratio of textile firms is derived by analogy.  
\begin{itemize}
	\item When $0<T\epsilon \le 1$, there exists a stable immovable point $x_t=0$, indicating that there is no set. That is, there does not exist a corresponding non-zero $x_t$ taking value.   
	\item When $1<T\epsilon \le 3$, $x_t$ has a locally stable structure with period $1$. If $T\epsilon$ takes any value in $(1, 3)$, $x_t$ takes the corresponding one between $(0, 0.664)$. This indicates that the innovation set $x_t$ is stable at that value, i.e. a $T\epsilon$ leads to only one innovation set $x_t$.      
	\item When $T\epsilon = 3$, the first bifurcation value $T\epsilon_1$ occurs and a forked bifurcation occurs. This will double the $x_t$ cycle. At this point the structure of $x_t$ goes from stable to unstable.    
	\item When $3<T\epsilon \le 3.4495$, a new locally stable structure of $x_t$ emerges, but at this point the period becomes $2$, i.e. a $T\epsilon$ parameter value will lead to two innovation sets $x_{t1}$ and $x_{t2}$. Over time, their innovation set bounces around $x_{t1}$ and $x_{t2}$, thus constituting two stable sequences of immobile point values that range from $0.664~0.85$ and $0.664~0.44$, respectively, as $T\epsilon$ varies. 
	\item When $T\epsilon = 0.4495$, a second bifurcation value $T\epsilon_2$ occurs and the Hough bifurcation occurs. $x_t$ has an unstable structure and thereafter the $x_t$ period multiplies to $4$.  
	\item When $3.4495<T\epsilon \le 3.5441$, the structure of $x_t$ changes from unstable to stable, and the period of $x_t$ is $4$. This indicates that one $T\epsilon$ parameter value will lead to four innovation sets, at which time $x_t$ has four sets of stable sequences of immobile point values, which range from 0.85-0.883, 0.85-0.82, 0.44-0.52 and 0.44-0.3367, respectively, with different values of $T\epsilon$.    
	\item When $T\epsilon=3.5441$ , a third bifurcation value $T\epsilon_3$ occurs, and the structure of $x_t$ is unstable, after which the period multiplies to $8$. 
	\item When $T\epsilon >3.5441$, the local structure of $x_t$ is stable and the period of $x_t$ is $8$. The rest of the process continues until $T\epsilon=3.5699=T\epsilon_{\infty}$, when the period of $x_t$ is infinite, i.e. chaos emerges. In other words, there is no relationship between the firm's innovation accumulation and the control parameters at this point.
\end{itemize}

\subsection{Sexual analysis of different parameter values}
Firstly, the values of $\alpha$, $\beta$, and $n$ are taken and defined as shown in Table \ref{tab1}, and all data are obtained from the findings of the "China Statistical Yearbook".

\begin{table}[H]
    \caption{\label{tab1} Variable definition and value}
    \centering
    \resizebox{\linewidth}{32pt}{
    \begin{tabular}{ccc}
        \hline
        Symbol & Description & Fetching method \\
        \hline
        $\alpha$ & Growth rate of investment in science and technology & Log difference of R\&D inputs \\
        $\beta$  & Proportion of technological content of output & Log difference in the number of active patents in the textile industry \\
        $n$     & Annual growth rate of the labour force & Log difference of the average number of workers employed \\
        $\epsilon$ & Government regulation parameters & By taking the product of $T\epsilon$ backwards or assuming that given \\
        \hline
    \end{tabular}}
\end{table}

From an econometric perspective, log-differencing helps to enhance the significance of the main regression to some extent. As a corollary, the significance of log-differencing the variables in this paper is to smooth out the effects of unreasonable outliers and reduce the "misleading" effect of episodic events on the overall judgment. In addition, as the average number of workers before 14 years could not be found directly, the calculation of business income/per capita business income was used, and the use of logarithmic difference to calculate the growth rate can also reduce the subtle influence of different quantiles on the conclusion. The log-difference results are shown in Table \ref{tab2} \footnote{All data are obtained from the China Statistical Yearbook, and are rounded to four decimal places after calculation by stata. The values of $\alpha$ range from 1.46\% to 47.44\%, $\beta$ from 9.58\% to 51.69\% and $n$ from -16.46\% to 4.79\%, so the values of $\alpha+\beta$ and $1+n$ are inferred from this.}.

\begin{table}[H]
	\caption{\label{tab2} Log-differential results}
	\centering
		\begin{tabular}{cccccc}
			\hline
			Year & $\alpha$ & $\beta$ & $n$ & $\alpha+\beta$ & $1+n$ \\
			\hline
			2009&/&/&/&/&/	\\
			2010&.2017&.5088&.0479&.7105&1.0479\\
			2011&.4744&.1671&-.0947&.6416&.9053\\
			2012&.0146&.1347&-.0566&.1494&.9434\\
			2013&.1382&.1418&-.0614&.2800&.9386\\
			2014&.1144&.5169&-.0653&.6313&.9347\\
			2015&.1558&.0958&-0.540&.2516&.9460\\
			2016&.0574&.2867&-.0627&.3441&.9373\\
			2017&.0584&.2688&-.1090&.3273&.8910\\
			2018&.0912&.2629&-.1646&.3541&.8354\\
			\hline
		\end{tabular}
\end{table}

Table \ref{tab2} shows that the values of $\alpha+\beta$ range from 14.94\% to 71.05\%, and the values of $1+n$ range from 83.54\% to 104.79\%. Considering that the value of the parameter $T$ in equation \ref{equ3} is proportional to $(\alpha+\beta)$ and inversely proportional to $(1+n)$, and that $T$ is related to the innovation accumulation $x_t$ of the textile enterprises in equation \ref{equ3}, the sexual attitude of $x_t$ can be analysed through this. To facilitate the calculation, the three fixed cases of the lower, middle and upper limits of these elements are taken separately, and only $\epsilon$ is adjusted to analyse the nature of $x_t$.

\begin{itemize}
	\item Lower bound case. $\alpha+\beta=0.1494$ , $1+n=0.8354$ . When $\epsilon=5.5917$, $T\epsilon=1.0000$ . From equation \ref{equ3} and Figure \ref{fig:fig3}, $x_t=0$. when $\epsilon=10$, $T\epsilon=1.7884$ and $x_t=0.4408$ . This means that at $(\alpha+\beta=0.1494)$, the degree of innovation accumulation achieved by textile firms remains zero when the parameter of government regulation of textile firms is 5.5917. When adjusted to the maximum value $\epsilon=10$, $T\epsilon=1.7884$ (satisfying $1<T\epsilon<3$), at which point the period of $x_t$ is $1$, the innovation intensification of textile enterprises is also only 44.08\%. At the same time, considering the nature of the enterprise cycle of China's existing textile enterprises and the characteristics of the industry, it is easy to see that the government's R\&D subsidies for the traditional textile industry are not significant enough. Therefore, it is difficult for the government's regulation parameter $\epsilon$ for the textile industry to reach a maximum value of $10$. That is, in the actual process, the innovation intensity of textile enterprises will be much less than 44.08\%.
	\item Middle bound case. This is calculated for the case $\alpha+\beta<0.2754$ , $1+n=0.9417$. When the government regulation parameter $\epsilon=10$, $T\epsilon=2.9245 (T\epsilon<3)$, it can be inferred that in this case, whatever the value of $\epsilon$ in the above range, the bifurcation does not occur and the structure is stable. Therefore, in this case, the government can increase the regulation of the innovation intensity of the textile enterprises, and can increase the policies and subsidies for them.
	\item Upper bound case. $\alpha+\beta=0.9913$ , $1+n=1.0479$ . When $\epsilon=1.0571$, $T\epsilon=1.0000$ , the innovation set is $0$; when $\epsilon<3.1659$, $T\epsilon<3=T\epsilon_1$ , the period of $x_t$ is $1$ and $x_t<0.3339$ , i.e. the innovation set is less than 33.39\%; when $3.1659<\epsilon<3.6465$, $3<T\epsilon<3.4495=T\epsilon_2$ ,the period of $x_t$ is 2; when When $3.6465<\epsilon<3.7465$, $3.4495<T\epsilon<3.5441=T\epsilon_3$ and the period of $x_t$ is 4. From the above, it can be seen that when $\epsilon_1=3.1659$, $\epsilon_2=3.6465$ and $\epsilon_3=3.7465$, there will be three bifurcations of $T\epsilon$, $T\epsilon_1$, $T\epsilon_2$ and $T\epsilon_3$. This means that $\epsilon_1$, $\epsilon_2$ and $\epsilon_3$ are dangerous values of strength that lead to a non-stationary state in the textile innovation accumulation model. As $\epsilon$ increases further, $T\epsilon$ also becomes larger until $\epsilon = 3.7737 = \epsilon_{\infty}$ when $T\epsilon = 3.5699 = T\epsilon_{\infty}$ and the period of $x_t$ is infinite, i.e. chaos will emerge. Clearly, $\epsilon=3.7737$ is the more dangerous regulatory parameter. At the upper value, $x_t$ takes values of 0.5270-0.6847 when the regulation parameter $\epsilon$ is kept between 2.1142-3.1713, which would make $T\epsilon$ take values between 2-3. This suggests that under a good state of affairs for firms, an appropriate increase in government regulation parameters would be beneficial to their innovation accumulation. It is also important to note that the government regulation parameter needs to be controlled below $3.7737$ in this case, otherwise textile firms will not be able to present a state with a stable cycle.	
\end{itemize}

\subsection{Further discussion}
According to the above discussion, we can streamline the results.	\\
\begin{itemize}
	\item At $\alpha+\beta \leq 0.1494$ and $n = -0.1646$, it is found that even with as much change as possible in the government regulation parameter $\epsilon$ for the textile industry, $T\epsilon$ cannot exceed 1.7884. Meanwhile, $x_t$ barely reaches 0.4408 and still does not reach 0.5, which means that increasing support for the textile industry at this point is very unreasonable and a waste of effective resources. This is because, with a declining textile workforce and limited development of its own, the textile industry is unable to use its resources effectively to generate the technological innovations it needs, and even greater support for the textile industry will not yield particularly effective results if the government does not manage to change the employment situation. In this case, the textile industry can only increase its $(\alpha+\beta)$ by investing in its own research and development in order to achieve an optimal value of $T\epsilon$ above 0.5.
	\item When $0.1494\leq \alpha+\beta \leq 0.2754$ and $-0.1646 \leq n \leq -0.0583$, it is found that: the reduction in the annual labour growth rate of the textile industry in such cases is not very serious, so that even if the government regulation parameter $\epsilon$ achieves its maximum value, $T\epsilon$ is only $2.9245 < 3 =T\epsilon_1$, at which point there is still no bifurcation and the period of $x_t$ is still 1. Therefore, the Continued and increased support for the textile industry can still have some amount of positive effect.
	\item When $0.2754 \leq \alpha+\beta\leq 0.9913$ and $-0.0583\leq n\leq 0.0479$, the situation becomes extremely complicated, but it is still possible to determine a reasonably appropriate parameter for the government's regulatory effort based on the values taken in different cases. It is worth noting that in such cases, the government's regulatory parameters should avoid the dangerous values of $\epsilon_1 = 3.1659$, $\epsilon_2 = 3.6465$, $\epsilon_3 = 3.7465$ and $\epsilon_{\infty} = 3.7737$. This is then adjusted in order to find the optimal control parameter $\epsilon$, which allows for a better accumulation of innovations in the textile industry.
\end{itemize}

\section{Conclusion}
Through the interpretation of chaos theory, the cumulative effect of technological innovation in the textile industry is explored from a non-linear perspective, and the level of innovation intensification is the level of estimation of innovation in the textile industry. It is easy to see that using the data from 2009-2018 as a base, the appropriate level of adjustment that the government should give under different circumstances can be well revealed. Specifically, this can be summarised as follows: 1). When an enterprise's workforce shows a more serious negative growth, the government should not directly support its technological innovation through a series of means such as technological subsidies, but should first start to address the situation of a declining workforce in the textile enterprise concerned. For example, the minimum wage for blue-collar workers should be increased. In order to effectively avoid a large loss of labour in the textile industry and thus maintain a good positive innovation accumulation effect in the textile industry. 2). When labour turnover in the textile industry is not obvious, the government should increase policy support for the textile industry to stimulate its potential power of technological innovation to ensure maximum technological capacity gains. 3). When the textile industry is essentially free of labour losses or even growth, the government, in judging the situation of the textile industry in order to provide appropriate regulation, should sufficiently avoid bifurcation points in order to ensure that the textile industry "goes to the next level" without losing its existing innovation accumulation and in order to stimulate its innovation surplus. The purpose of avoiding bifurcation points is to avoid the uncertainty of direction leading to a decrease rather than an increase in the innovation intensity of the textile industry, but care should always be taken to avoid chaos.	\\
At the same time, there is a developable period of ongoing research in the article. The model in this paper only establishes the variable $\epsilon$, the parameter of government regulation intensity, but does not find a very suitable quantitative criterion that can be used to measure this parameter. It is hoped that future research by scholars will further explore the quantitative indicators concerning the parameter of government regulation intensity affecting the textile industry. It is hoped that future research will further explore the quantitative indicators that affect the strength of government regulation in the textile industry and form a complete macro chaos model that will help the government to regulate the textile industry.

\section*{Statement}
\begin{itemize}
	\item This paper was completed at the end of my undergraduate degree, so the overall idea of writing is underdeveloped.
	\item This article is not sufficiently rigorous and provides only a brief analysis, which may not even be considered an analysis.	
\end{itemize}

\bibliographystyle{unsrt}  
\bibliography{references}

\end{document}